\documentclass[prb,aps,graphicx,preprint]{revtex4-1}
\usepackage{graphicx}
\usepackage{amssymb}
\usepackage{color}
\begin{document}

\title{Magnetoelastic coupling induced magnetic anisotropy in Co$_2$(Fe/Mn)Si thin films}

\author{Himanshu Pandey,$^1$ P. K. Rout,$^1$ Anupam,$^1$ P. C. Joshi,$^1$ Z. Hossain,$^1$ and
R. C. Budhani$^{1,2\ast}$}
\affiliation{$^1$Condensed Matter-Low
Dimensional Systems Laboratory, Department of Physics,
Indian Institute of Technology Kanpur, Kanpur-208016, India\\
$^2$CSIR-National Physical Laboratory, New Delhi-110012, India}
\email{rcb@iitk.ac.in, rcb@nplindia.org}
\date{\today}

\begin{abstract}
\baselineskip 0.8cm

The influence of epitaxial strain on uniaxial magnetic anisotropy
of Co$_{2}$FeSi (CFS) and Co$_{2}$MnSi (CMS) Heusler alloy thin
films grown on (001) SrTiO$_3$ (STO) and MgO is reported. The
in-plane biaxial strain is susceptible to tune by varying the
thickness of the films on STO, while on MgO the films show
in-plane easy axis for magnetization (\overrightarrow{M})
irrespective of their thickness. A variational analysis of
magnetic free energy functional within the Stoner-Wohlfarth
coherent rotation model with out-of-plane uniaxial anisotropy for
the films on STO showed the presence of magnetoelastic anisotropy
with magnetostriction constant $\approx$
(12.22$\pm$0.07)$\times$10$^{-6}$ and
(2.02$\pm$0.06)$\times$10$^{-6}$, in addition to intrinsic
magnetocrystalline anisotropy $\approx$ -1.72$\times$10$^{6}$
erg/cm$^{3}$ and -3.94$\times$10$^{6}$ erg/cm$^{3}$ for CFS and
CMS, respectively. The single-domain phase diagram reveals a
gradual transition from in-plane to out-of-plane orientation of
magnetization with the decreasing film thickness. A maximum
canting angle of 41.5$^{\circ}$ with respect to film plane is
predicted for the magnetization of the thinnest (12 nm) CFS film
on STO. The distinct behaviour of \overrightarrow{M} in the films
with lower thickness on STO is attributed to strain-induced
tetragonal distortion.
\end{abstract}

\maketitle \baselineskip 0.8cm

\section{Introduction}
The Heusler alloys have taken the center stage as spintronics
materials due to their high degree of spin polarization, high
Curie temperature, and low magnetic damping.\cite{Katsnelson,Graf}
By tuning the magnetic parameters such as coercivity, anisotropy,
exchange interactions and damping processes, one can suitably
tailor these materials for magnetic random access memory, magnetic
logics, spin-transistors, and related potential applications.
However, in most of such applications the magnetic alloy has to be
in a thin film form in which its magnetic characteristics can be
significantly different due to film thickness, crystallographic
orientation, growth related strains and interfacial reactions. One
such characteristics is magnetic anisotropy, which should be large
for magnetic storage applications, and which also determines the
magnetization reversal processes in magnetic switching devices.
Till now, a large number of full-Heusler alloy thin films have
been grown on various substrates. Some examples of this are
Co$_2$MnGe on GaAs\cite{Yang} and Al$_2$O$_3$,\cite{Belmeguenai}
Co$_2$MnSi on GaAs,\cite{Wang} MgO,\cite{Him,Him1,Bosu} and
Al$_2$O$_3$,\cite{LJSingh} Co$_2$FeAl$_{0.5}$Si$_{0.5}$ on
MgO,\cite{Trudel} Co$_2$FeSi on GaAs,\cite{Hashimoto,Jenichen}
Al$_2$O$_3$,\cite{Schneider} and MgO\cite{Schneider} as well as on
SrTiO$_3$ (STO).\cite{Anupam,Prasanna,Him2} While the substrate
lattice parameter, growth, thermal annealing condition and film
thickness in these cases vary significantly, the effect of such
condition on magnetic anisotropy of the films is seldom addressed.
In Heusler alloys films, one expects a four-fold anisotropy due to
the cubic symmetry of the unit cell, while in-plane uniaxial
anisotropy has also been observed for the case of Co$_2$FeSi grown
on GaAs.\cite{Hashimoto} The presence of additional uniaxial
anisotropy has resulted in multistep magnetization switching in
some Heusler alloy films.\cite{Yang,Wang} Moreover, Gabor
\textit{et al.} have shown that Co$_{2}$FeAl films can have three
types of in-plane anisotropies, namely biaxial (fourfold cubic
anisotropy) and two uniaxial anisotropies parallel to the biaxial
easy and hard axes.\cite{Gabor} In some cases, stripe domains have
also been seen due to magnetic frustration between two
energetically equivalent easy axis.\cite{Liu} The interface
between the film and substrate also affects the orientation of
magnetization significantly. For example, the out-of-plane
magnetic easy axis in Co$_{2}$FeAl films on Cr-buffered MgO
substrate seemed to be induced by the interfacial anisotropy which
appears after annealing the films in the presence of magnetic
field applied along out-of-plane direction.\cite{Wen}\\

The magnetic anisotropy in thin films originates from fundamental
factors such as the spin-orbit interaction in the material which
controls magnetocrystalline anisotropy and/or due to growth
related strain. Any change in the lattice via strain will change
the distances between the magnetic atoms and alter the interaction
energy, which decides the magnetoelastic anisotropy. The strain
therefore becomes a tuning parameter for magnetic anisotropy and
can be varied by a choice of substrates of different lattice
parameter or films of varied thickness. A consequence of the
strain related anisotropy is the rotation of magnetic easy axes
from in-plane to out-of-plane configuration or vice versa. While a
strain dependence of in-plane anisotropy has been reported for
Co$_2$FeAl/MgO thin films,\cite{Gabor} to the best of our
knowledge, strain driven out-of-plane anisotropy has not been
reported for Heusler alloy films. Here we report a detailed study
of the magnetic anisotropy of Co$_2$(Fe/Mn)Si [CF(M)S] films of
various thickness deposited on (001) MgO and (001) STO crystals.
The in-plane biaxial strain was gradually varied from compressive
(for the films on STO) to tensile (for the films on MgO) by
depositing the films of different thickness. We have specifically
focussed on the strain dependence of out-of-plane uniaxial
magnetic anisotropy in CF(M)S/STO films and established how the
strain induced magnetic anisotropy affects the direction of
magnetization (\overrightarrow{M}). It is seen that the tuning of
magnetoelastic coupling by varying the film thickness results in
the rotation of the magnetization vector towards out-of-plane
direction as the film thickness is lowered.

\section{Experimental details}
We have previously demonstrated that the CF(M)S films on SrTiO$_3$
and MgO processed at 600$^{\circ}\mathrm{C}$ have better
crystalline quality as compared to those annealed at lower
temperatures.\cite{Anupam,Him,Him1} Therefore, for studies of
anisotropy reported here, we mainly concentrate on the films
processed at 600$^{\circ}\mathrm{C}$. The cubic lattice parameter
($a_{bulk} \approx$ 0.5656 nm) of CF(M)S matches quite well with
the face diagonal ($\sqrt2a_{sub}$) of (001) STO and MgO. The
lattice misfit [$\epsilon$ = $(a-\sqrt2a_{sub} )/\sqrt2a_{sub}$]
of CM(F)S with STO and MgO lies within 6$\%$. Taking advantage of
close matching of the lattice parameters, we have prepared a
series of CF(M)S thin films of various thickness ($\textit{t}$ =
5-100 nm) epitaxially on (001) STO and MgO using pulsed laser
deposition technique. The details of thin film preparation are
described in our earlier reports.\cite{Him,Him1,Anupam,Prasanna}
The structural characterization of the films has been done by
X-ray diffraction (PANalytical X$'$Pert PRO X-ray diffractometer)
in $\theta$-2$\theta$, $\omega$, $\varphi$, and grazing incidence
X-ray diffraction (GIXRD) modes. The magnetic measurements were
performed in a vibrating sample magnetometer (EV7 VSM) at room
temperature.

\section{Results and discussion}
\subsection{Structural characterization}

\begin{figure}[t!]
\centering \vskip -0cm \abovecaptionskip 0.1cm
\includegraphics [width=12cm]{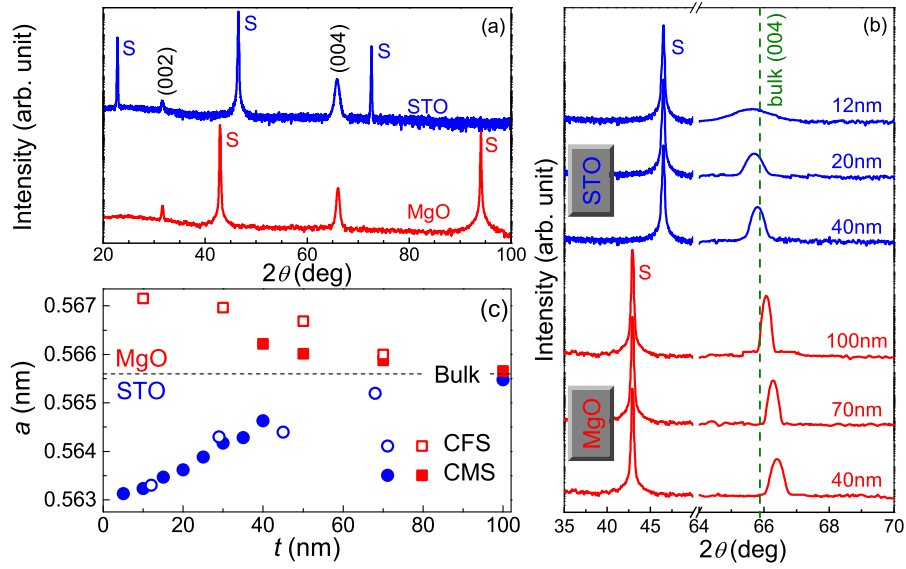}
\caption{\label{Fig 1} (Color online) (a) The $\theta-2\theta$
X-ray diffraction profiles of 40 nm thick CMS films grown on STO
and MgO. (b) The small range $\theta-2\theta$ scan about (004)
peak of CMS films on STO and MgO with thickness. The Bragg
reflections from (002) planes of the substrates are also shown.
The dashed line shows the position of 2$\theta$ value
corresponding to (004) peak of bulk CMS. (c) The in-plane ($a$)
lattice parameter as a function of film thickness $t$ for CMS
(filled symbols) and CFS (empty symbols). The cubic lattice
parameter of bulk CMS is markrd by the dotted line.}
\end{figure}

The $\theta-2\theta$ X-ray diffraction reveals (00$l$) oriented
growth of CF(M)S films on STO and MgO [Fig. 1(a)]. Further
evidence of (00$l$) texturing is provided by the rocking curves
about (004) reflection. The full width at half maximum of these
films are less than 1.9$^\circ$, which corresponds to a
crystallite size of 5 nm.\cite{Prasanna} Moreover, the $\varphi$
scans confirm the epitaxial growth of the films with the relation
[100] CF(M)S $\parallel$ [110] STO or MgO. The presence of (111)
superlattice, which governs the ordering of the Mn(or Fe) and Si
sublattices, and (022) fundamental diffraction line, which
confirms the presence of \textit{$L2_1$} ordering in the films,
are two important indicators of the structural ordering in the
films. From GIXRD measurements, we infer the degree of ordering in
the films to be more than 85$\%$. Figure 1(b) shows the
$\theta-2\theta$ scan about (004) peak for films of various
thickness on MgO and STO substrates. A clear shift of the Bragg
reflections towards higher (lower) scattering angle ($2\theta$) is
seen for the films grown on MgO (STO) as the thickness is reduced.
The out-of-plane lattice parameter ($c$) obtained from these scans
decreases (increases) for the films grown on STO (MgO) with the
increasing $t$. This can be understood in terms of the strain
induced in the films due to lattice misfit. The positive misfit
value for STO ($\epsilon$ = 2.4$\%$) results in in-plane
compressive strains, which decreases the in-plane lattice
parameter ($a$) as verified by off-axis $\theta-2\theta$ scans
about (022) peak. Assuming the volume ($a^2c$) preserving
distortion, we expect an increase in $c$ with decreasing $t$ for
the films grown on STO. The films with lower thickness experience
a relatively strong tetragonal distortion. As the film thickness
increases, the distortion relax by formation of misfit
dislocations. With increase in thickness, the in-plane strain
$\epsilon_{xx}$ [=($a_{film}-a_{bulk}$)/$a_{bulk}$] approaches
zero as seen in Fig. 1(c). We observe that the thinnest film ($t
=$ 5 nm) on STO is under highest biaxial compressive strain of
$\epsilon_{xx} = \epsilon_{yy}$ = -0.44 \% while the thicker films
undergo partial strain relaxation with 100 nm film attaining bulk
values. Similarly, the tensile strain in the Heusler alloy films
on MgO disappears on increasing their thickness.

\begin{figure}[t!]
\centering \vskip -0cm \abovecaptionskip -0.5cm
\includegraphics [width=12cm]{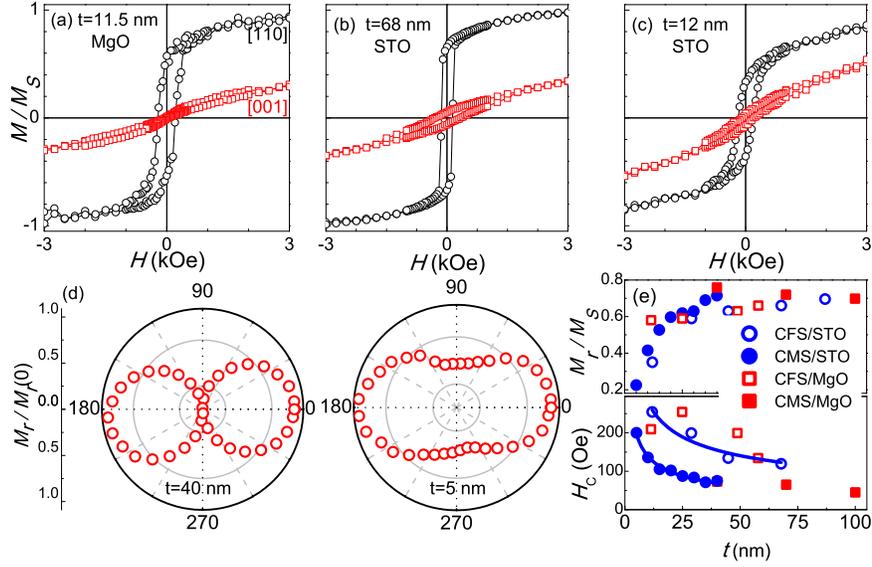}
\caption{\label{Fig 2} (Color online) The magnetic hysteresis
loops measured along [110] and [001] directions at room
temperature of (a) CFS (11.5 nm)/MgO, as well as (b) 68 nm and (c)
12 nm thick CFS/STO films. (d) Polar plot of $M_{r}/M_{r}$(0) for
5 nm and 40 nm thick CMS/STO film at a step of 5$^{\circ}$. Here
$M_{r}$(0) is the $M_{r}$ at $\theta=0^{\circ}$. (e) The upper
panel shows the thickness dependence of $M_{r}/M_{S}$, where
$M_{S}$ is the saturation moments. The Slater-Pauling formula
predicts a $M_S$ of 5$\mu_B$ and 6$\mu_B$ for CMS and CFS films,
respectively.\cite{Galanakis} We have used the experimental values
of $M_S$, which are in reasonable agreement with the
theory.\cite{Prasanna} The lower panel shows the $H_C$ as a
function of $t$ along with the fits (solid line) according to the
relation: $H_C \propto t^{-n}$.}
\end{figure}

\subsection{Magnetization}
We first discuss the behavior of magnetic hysteresis loops
[$M(H)$] for in-plane (along [110]) and out-of-plane (along [001])
field configurations [See Fig. 2(a-c)]. The hysteresis loops of
the films on MgO clearly show an in-plane easy axis for
magnetization ($\overrightarrow{M}$) as revealed by the squareness
of the loop in Fig. 2(a). This result is the same for thicker CFS
films on STO [Fig. 2(b)]. However, for our thinnest film on STO,
we observe a significantly higher out-of-plane magnetization,
which suggests the possibility of tilted $\overrightarrow{M}$ with
respect to the film plane. Figure 2(d) shows the remanent
magnetization ($M_r$) at different angles ($\theta$) of the field
with respect to the film plane, which looks like a dumbbell with
two lobes almost separated from each other for CMS(40 nm)/STO
film. Clearly, the $M_r$/$M_r (0)$ is maximum for
$\theta=0^{\circ}$ and 180$^{\circ}$ (in-plane directions) while
it is almost zero at 90$^{\circ}$ and 270$^{\circ}$ (out-of-plane
directions). This observation confirms the presence of in-plane
easy axis for thicker films on STO. However, in the case of CMS(5
nm)/STO film, two lobes are joined and thus the $M_r$ is
substantially higher for $\theta =$ 90$^{\circ}$ and
270$^{\circ}$. This suggests a canted easy axis instead of an
in-plane one as observed in thicker films. We believe that the
substrate-film interface plays an important role in tilting the
magnetization away from the film plane. The upper panel in Fig.
2(e) shows the thickness dependence of the squareness
($M_{r}/M_S$) of magnetization extracted from in-plane M(H) loops.
In case of films on MgO, it remains almost constant whereas, for
the films on STO, we notice a gradual decline in $M_r$/$M_S$ with
decreasing thickness, which indicates the deviation of easy axis
from the film plane. Although the lowest observed value ($\approx$
0.2 for 5 nm film) does not point towards a distinct out-of-plane
easy axis, it certainly indicates some canting of
$\overrightarrow{M}$ away from the film plane.\\

The coercivity of a material is the principal property related to
the rate of change of magnetic relaxation between the remanent and
demagnetized states. At absolute zero, it measures the barrier
height that is required by magnetic moments to overcome the
demagnetized state. The variation of the coercivity ($H_C$) of the
films with thickness is plotted in the lower panel of Fig. 2(e).
We observe that $H_C$ decreases gradually with increasing
thickness in all cases. This may be attributed to a lowering of
defect concentration due to enhancing crystalline quality or due
to lowering of strain in thicker films. Moreover, the reduction of
$H_C$ can also be due to the changes in the grain size and the
surface roughness of the film with its thickness or related to the
fact that the film thickness decreases to a point where the domain
wall thickness becomes comparable to the film thickness. The $H_C$
follows a power law type dependence on $t$ of the form: $H_C
\propto$ t$^{-n}$ with $n = 0.50\pm 0.02$ and $0.41\pm0.17$ for
CMS and CFS films on STO, respectively. The value of $n$ depends
on the deposition conditions and the choice of ferromagnet, and
can have values from -0.3 to -1.5.\cite{Neel,Tolman,Wolf}

We have carried out an analysis of the hard axis magnetization
loops in the framework of Stoner-Wohlfarth formalism.\cite{Stoner}
The total magnetic free energy ($E$) of the film in tetragonal
symmetry can be expressed as:

\begin{eqnarray}
E = K_1 m_z^2+K_2 m_z^4+K_3 m_x^2 m_y^2-\overrightarrow{M} \cdot \overrightarrow{H}+2\pi M_s^2 m_z^2
\end{eqnarray}

where $K_{1}$ and $K_{2}$ are second and fourth order uniaxial
anisotropy constants, respectively while $K_{3}$ is in-plane
biaxial anisotropy constant. The $m_{x,y,z}$ are the direction
cosines of the magnetization vector $\overrightarrow{M}$. The
fourth term of Eq. (1) is the Zeeman energy and the last term
represents the thin film demagnetization energy. For out-of-plane
field hysteresis loop \emph{i.e}. when $\overrightarrow{H}$ is
applied along [001], $\overrightarrow{M}$ will rotate from the
[110] (in-plane easy axis) to [001] direction and thus the term
$K_{3}{m_{x}}^{2}{m_{y}}^{2}$ is always zero. The minimization of
total magnetic free energy for an out-of-plane field yields the
equilibrium magnetization $M$ in the field direction given by the
relation:

\begin{eqnarray}
H = \left[ {\frac{{2K_1 }}{{M_S^2 }} + 4\pi } \right]M +
\frac{{4K_2 }}{{M_S^4 }}M^3
\end{eqnarray}

The values of $K_{1}$ and $K_{2}$ can be obtained by fitting Eq.
(2) to the hysteresis loops. The inset of Fig. 3(a) shows the plot
of $H/M$ vs. $M^2$ for 68 nm thick CFS/STO film. The intercept and
slope of the linear fit yield $K_{1}$ and $K_{2}$, respectively.
The deviation in upper part of the curve from the linearity occurs
as $M$ approaches saturation, while the deviation at lower $M$ can
be attributed to magnetic domain effects.\cite{Wang2} All the
films on MgO show in-plane easy axis without any substantial
change in $M$($H$) loops with thickness. So the determination of
anisotropy coefficients for these films will not be reliable.
While we observed clear change in $M$($H$) loops for the films on
STO with varying thickness. Hence we will only focuss on the later
films in order to gain further insight of the magnetic state.

\begin{figure}[h]
\centering \vskip -0cm \abovecaptionskip -0.5cm
\includegraphics [width=12cm]{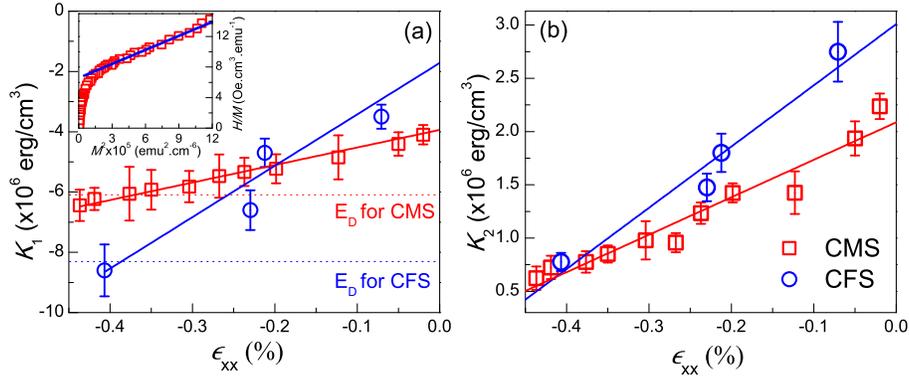}
\caption{\label{Fig 3} (Color online) (a) The second order uniaxial anisotropy
constant ($K_{1}$) as a function strain ($\epsilon_{xx}$) with the
linear fits (solid lines). The $\epsilon_{xx}$ has been calculated using the
values of $a$ mentioned in Fig. 1(c). The dotted lines show
demagnetization energy ($E_D$) for CF(M)S. The inset shows the
plot of $H/M$ vs $M^2$ for CFS (68 nm)/STO film along with the linear
fit (solid line) given by Eq. (2).(b) The fourth order uniaxial anisotropy
constant ($K_{2}$) as a function of $\epsilon_{xx}$ with the
linear fits (solid lines).}
\end{figure}
Figure 3 shows the values of $K_1$ and $K_2$ deduced from Eq. (2)
for CF(M)S/STO films as a function of $\epsilon_{xx}$. We clearly
observe a monotonic increase in anisotropies with the increasing
strain. Moreover, the values of $K_1$ are quite similar to
previously reported values.\cite{Him} The $K_1$ is connected to
$\epsilon_{xx}$ through the magnetoelastic coupling parameters and
can be expressed as $K_1$ = $K_{mc} + 3\lambda
\sigma_{xx}$/2.\cite{Culity} The first term represents the strain
independent magnetic anisotropy, commonly known as
"magnetocrystalline anisotropy", which originates from the
inherent crystal structure of ferromagnet.\cite{Culity} The linear
fits to $K_1 (\epsilon_{xx})$ data yield $K_{mc} \approx$
-1.72$\times$10$^{6}$ erg/cm$^{3}$ and -3.94$\times$10$^{6}$
erg/cm$^{3}$ for CFS and CMS, respectively [See Fig. 3(a)]. The
second term is purely related to the strain induced anisotropy,
which depends linearly on stress and the magnetostriction constant
$\lambda$. The stress can be represented as $\sigma_{xx} = Y
\epsilon_{xx}$, where the Young$'$s modulus ($Y$) can be expressed
in terms of elastic stiffness constants ($C_{11}$ and $C_{12}$) as
follows: $Y = (C_{11}-C_{12})(C_{11}+2C_{12})/(C_{11}+C_{12}
)$.\cite{Ledbetter} Assuming theoretical values of
$C$'s,\cite{Chen} we find $Y \approx$ 93 GPa for CFS and 192 GPa
for CMS. Using these values, the linear fits to
$K_1(\epsilon_{xx})$ data yield $\lambda \approx$
(12.22$\pm$0.07)$\times$10$^{-6}$ and
(2.02$\pm$0.06)$\times$10$^{-6}$ for CFS and CMS, respectively. To
the best of our knowledge, we are unaware of any values of
$\lambda$ and $K_{mc}$ for these compounds reported in literature.
The values of $\lambda$ are comparable to the reported value of
$\sim$ 15$\times$10$^{-6}$ for another Heusler alloy
Co$_2$MnAl\cite{Qiu} while $\lambda$ is of the order of 10$^{-5}$
for half metallic manganites.\cite{Donnell,Lofland}

\begin{figure}[h]
\centering \vskip -0cm \abovecaptionskip -0.1cm
\includegraphics [width=8cm]{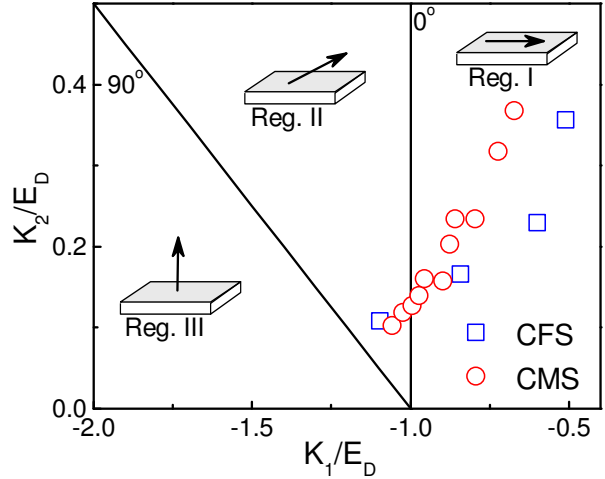}
\caption{\label{Fig 4} (Color online) The single-domain magnetic
phase diagram demonstrating different stable magnetic states,
namely in-plane (Region I), canted (Region II), and out-of-plane
(Region III) state of magnetization. The symbols are the
experimental data.}
\end{figure}

Our expression for $K_1$ in case of biaxial stress ($\sigma_{xx} =
\sigma_{yy}, \sigma_{zz} = 0$) is same as the expression for
uniaxial stress ($\sigma_{xx} \neq 0$, $\sigma_{yy} = \sigma_{zz}
= 0$) induced anisotropy, i.e. $K = 3\lambda \sigma_{xx} / 2$. But
these two cases are fundamentally different. In the former
scenario, a uniaxial anisotropy is induced perpendicular to the
plane (along $z$-axis) while for latter case the uniaxial
anisotropy is along the direction of applied stress (along
$x$-axis). The other anisotropy constant $K_2$ also shows a linear
dependence with $\epsilon_{xx}$ as shown in Fig. 3(b). Such linear
relation has been predicted for a cubic system under biaxial
strain and experimentally verified for Cu-Ni
systems.\cite{Handley} Similar to the case for $K_1$, we observe a
substantial contribution to $K_2$ coming from magnetocrystalline
origin in addition to the magnetoelastic couplings.

The direction of magnetic easy axis depends sensitively on
anisotropy energy dependent on $K_1$ and $K_2$ and the
demagnetization energy $E_D$ (=2$\pi M_s ^2$). Only consideration
of second-order angular term gives an out-of-plane magnetization
state for $K_1$/$E_D$ $<$ -1 while $\overrightarrow{M}$ becomes
in-plane for -1 $<$ $K_1$/$E_D$. However, the fourth order
anisotropy term introduces the canting states of
$\overrightarrow{M}$ allowing a gradual transition between the
in-plane and out-of-plane states.\cite{Handley,Gyanendra} Figure 4
shows the general single-domain magnetic phase diagram for a
system with free energy given by Eq. (1) in zero magnetic field
assuming a coherent rotation of magnetization. The films whose
anisotropy data lie in Region II have canted magnetization states,
where the canting angle $\theta_c$ (the angle between
$\overrightarrow{M}$ and film plane) can be obtained from the
relation:\cite{Handley} $\sin^2\theta_c = -(K_1+E_D)/2K_2$. The
CFS (12 nm)/STO film has $\theta _c$ = 41.5$^{\circ}$ while the
angles for 5 nm and 10 nm thick CMS/STO films are 31.8$^{\circ}$
and 17.9$^{\circ}$, respectively. The data for thicker films fall
into Region I, which suggests that easy axis of magnetization is
in-plane. Clearly, it can be inferred that easy axis changes from
in-plane to canted orientation with increasing compressive strain.
Hence, there is a possibility to get the perpendicular magnetic
anisotropy in case of films with higher strain. This can be
achieved either by lowering the film thickness or choosing a
substrate with a larger positive misfit.

\section{Summary}
We have presented a study to correlate the crystallographic
structure and the magnetic state of Co$_{2}$FeSi and Co$_{2}$MnSi
films on (001) STO and MgO substrates. The films on STO are under
in-plane biaxial compressive strain while a tensile strain is
observed in the films on MgO. The strain gradually relaxes with
increasing film thickness. The hysteresis loops clearly show an
in-plane easy axis for all the films on MgO, however, for the
films on STO, the out-of-plane component of magnetization
increases with decreasing thickness. The analysis of magnetic free
energy functional within the Stoner-Wohlfarth coherent rotation
model with out-of-plane uniaxial anisotropy predicts a canted
magnetization state for the films on STO, which gradually moves
towards in-plane state with increasing thickness in a
single-domain magnetic phase space. The uniaxial anisotropy terms
have two distinct contributions; first one is intrinsic
magnetocrystalline anisotropy, which is strain independent and the
other one is magnetoelastic anisotropy. We have extracted various
anisotropy terms ($\sim$ 10$^6$ erg/cm$^3$) and magnetostriction
constants $\sim$ 10$^{-6}$ of Co$_{2}$FeSi and Co$_{2}$MnSi for
the first time. We also predict maximum canting angles of
41.5$^{\circ}$ and 31.8$^{\circ}$ for Co$_2$FeSi (12 nm) and
Co$_2$MnSi (5 nm) on STO, respectively. These results prove that
the epitaxial strain is a useful parameter to tailor the magnetic
anisotropy in thin film of Heusler alloys, which could lead to the
realization of out-of-plane magnetic anisotropy on oxide
substrates for fabrication of memory devices.

\section{Acknowledgements}
The authors thank M. Shivakumar for his help in magnetization
measurements. H.P. and P.K.R. acknowledge financial support from
Indian Institute of Technology Kanpur and Council for Scientific
and Industrial Research (CSIR), Government of India. This work is
partly supported by CSIR, New Delhi [Grant No.
80(0080)/12/EMR-II]. R.C.B. acknowledges J. C. Bose Fellowship of
the Department of Science and Technology, Government of India.

\end{document}